\documentclass[11pt]{article}
\usepackage{latexsym}
\usepackage{amssymb,cite,epsfig}

\def\be{\begin{equation}}
\def\ee{\end{equation}}
\def\bea{\begin{eqnarray}}
\def\eea{\end{eqnarray}}
\def\ba{\begin{array}}
\def\ea{\end{array}}
\def\nn{\nonumber \\}

\newcommand{\ul}{\underline}

\newcommand{\Gphi}{\Gamma^{\ul{\varphi}}}
\newcommand{\hGphi}{\hat\Gamma^{\ul{\varphi}}}

\begin{document}
\begin{flushright}
IFUM-760-FT\\
\end{flushright}
\vspace{.5truecm}

\centerline{\Huge  The explicit form of the effective action}
\centerline{\Huge  for F1 and D-branes }
\vspace{1truecm}

\begin{center}
\renewcommand{\thefootnote}{\fnsymbol{footnote}}
 {\Large Donald~Marolf$^{1}$\footnote{marolf@physics.ucsb.edu},
 Luca~Martucci$^{2,3}$\footnote{luca.martucci@mi.infn.it}
 and Pedro~J.~Silva$^{2,3}$\footnote{pedro.silva@mi.infn.it}}\\
\renewcommand{\thefootnote}{\arabic{footnote}}
\setcounter{footnote}{0}
\vspace{.5truecm} {\small \it $^1$ Physics Department, \\ 
UCSB, Santa Barbara CA 93106, USA.

\vspace*{0.5cm}

$^2$ Dipartimento di Fisica dell'Universit\`a di Milano,\\
Via Celoria 16, I-20133 Milano, Italy\\

\vspace*{0.5cm}

$^3$ INFN, Sezione di Milano,\\
Via Celoria 16,
I-20133 Milano, Italy\\
}

\end{center}

\vspace{1.5truecm}

\centerline{\bf ABSTRACT} \vspace{.5truecm} \noindent In this work we consider the full interacting effective actions for fundamental strings and D-branes in arbitrary bosonic type II supergravity backgrounds. The explicit form of these actions is given in terms of component fields, up to second order in the fermions. The results take a compact form exhibiting $\kappa$-symmetry, as well as supersymmetry in a background with Killing spinors. Also we give the explicit transformation rules for these symmetries in all cases.
\newpage

\section{Normal coordinate expansion in M-theory}

This work is based on the talk given in the RTN workshop 2003  ``The quantum structure of spacetime and the geometric nature of fundamental interactions'' in  Copenhagen, and contains a summary of the work that appeared in the three papers \cite{ms,mms1,mms2}. In these articles we were interested in studying the world-volume theory of various branes of M-theory and in particular on their fermionic sector and supersymmetries. We obtained the corresponding actions in terms of background component fields using the so-called ``normal coordinate expansion'' \cite{nc1,nc2}. We have worked out these expansions up to second order in the fermionic coordinates. The original strategy used to obtain all D-brane and fundamental string actions \cite{Bergshoeff:2001dc,Bergshoeff:1996tu,Cederwall:1996ri,Aganagic:1996pe} was to begin with the 11D supermembrane \cite{Bergshoeff:1987cm} and then, by single dimensional reduction (double dimensional reduction) to 10D, obtain the D2-brane (the fundamental string of type IIA). From here, by means of  the correct application of the pertinent form of t-duality rules \cite{Bergshoeff:1995as,has}, all the D-branes of type IIA/B (and the fundamental string of type IIB) were found.
In the following, the conventions and definitions used can be found in the three articles \cite{ms,mms1,mms2}. Nevertheless, we follow mainly the same conventions of \cite{Howe,Cremmer} for 11D supergravity and \cite{has,antoine} for 10D supergravity.

In a superspace formalism, the supercoordinates $z^M$ decompose into bosonic coordinates $x^m$ and fermionic coordinates $\theta^\mu$. Here we also introduce a similar decomposition for tangent space vectors $y^A$, with $A=(a,\alpha)$:
\bea
&&z^M=(x^m,\theta^\mu), \nn
&&y^A=(y^a,y^\alpha).
\eea
The normal coordinate expansion is a method based on the definition of normal
coordinates in a neighborhood of a given point $z^M$ of superspace.  
The idea is to parameterize the neighboring points by the tangent vectors
along the geodesics joining these points to the origin. Denoting the coordinates at
neighboring points by $Z^M$ and the tangent vectors by $y^A$, we have
\be
Z^M=z^M+\Sigma^M(y)\;,
\ee
where the explicit form of $\Sigma^M(y)$ is found iteratively by solving the
geodesic equation. Tensors at the point $Z^M$ may be compared with those at $z^M$ by parallel transport.  In this
sense, the change in a general tensor under an infinitesimal displacement $y^A$ is
\be
\delta T = y^A\nabla_A T\; .
\label{exp1}
\ee
Finite transport is obtained by iteration. In this way we may consider the corresponding expansion  in the operator $\delta$ for any tensor in superspace. For example,
consider the vielbein $E_M^A$
\be
E_M^{\;\;\;A}(Z)=E_M^{\;\;\;A}(z)+ \delta E_M^{\;\;\;A}(z)+ {1\over2}\delta^2E_M^{\;\;\;A}(z)+\ldots
\ee
In particular one finds the following fundamental relations by means of which one can expand any tensor iteratively to any order (see for example \cite{gk1}),
\bea
&&\delta y=0\;, \nn
&&\delta E^A= \nabla y^A + y^CE^BT_{BC}^{\;\;\;\;A}\;, \nn
&&\delta \nabla y^A = -y^BE^Cy^DR_{DCB}^{\;\;\;\;\;\;\;\;A}\;,
\eea
where $T$ is the torsion and $R$ the Riemann tensor.

In our case, since our starting point is the M2-brane, we are interested in an expansion up to second order around a bosonic background of 11D supergravity. We therefore set $z^M=(x^m,0)$ and $y^A=(0,y^\alpha)$. By means of the 11D superconstraints \cite{Howe,Cremmer}, one finds the following formulas:
\bea
&&\delta E^a =0, \nn
&&\delta E^\alpha = Dy^\alpha +y^\beta e^b T_{b\beta}^{\;\;\;\alpha}\;, \nn
&&\delta^2 E^a =-i(y^\alpha\Gamma^a_{\;\;\alpha\beta}Dy^\beta+y^\alpha y^\beta T_{\;\beta}^{\;\;\;\gamma}\Gamma^a_{\;\;\gamma\alpha})\;, \nn
&&\delta^2 E^\alpha = 0\;.
\label{exp2}
\eea
The M2-brane supersymmetric action is,
\be
S=-T_{M2}\int{d^3\xi\sqrt{\det(-\textbf{G})}}-
{T_{M2}\over 6}\int{d^3\xi\varepsilon^{ijk}\textbf{A}_{kji}},
\ee
where $i=(0,1,2)$, $T_{M2}=(4\pi^2 l_p^{3})^{-1}$, $l_p$ is the 11D Planck length, ($\textbf{G}$,$\textbf{A}$) are the pull-back to the supermembrane of the metric and 3-form super-fields of N=1 11D supergravity. At this point, in order to obtain the expanded M2-brane action in component fields, we simply take the M2-brane supersymmetric action, perform substitutions using equations (\ref{exp1},\ref{exp2}) and discard terms above second order in fermions. The resulting form of the action is
\bea
S_{M2}=-T_{M2}\int{d^3\xi\sqrt{-\det(G)}}-
{T_{M2}\over 6}\int{d^3\xi\varepsilon^{ijk}A_{kji}}+\cr
+\frac{iT_{M2}}{2}\int d^3\xi\sqrt{-G}\bar{y}(1-\Gamma_{M2})\Gamma^i\tilde{D}_iy\ .\label{m2}
\eea
where ($G$,$A$) are the bosonic part of the corresponding superfields, $y$ is a 11D Majorana spinor, $\label{gammaM2} \Gamma_{M2}=\hbox{${1\over 3!\sqrt{-G}}$}\epsilon^{ijk}\Gamma_{ijk}$, $\label{gravitino11d}
\tilde D_m=\nabla_m -\frac{1}{288}(\Gamma_m{}^{pqrs}-8\delta_m^p\Gamma^{qrs})H_{pqrs}$, $H=dA$, ($m,n..$) are 11D space-time indices and $\tilde D_i$ is the pull-back of $\tilde D_m$.

$\kappa$-symmetry transformation rules relevant for the above action (derived using normal coordinate expansion methods on the superfield form of the $\kappa$-symmetry transformation) are 
\bea
\label{km2} \delta_{\kappa}y&=&(1+\Gamma_{M2})\kappa+O(y^2)\ , \cr
\delta_{\kappa}x^{m}&=&\frac i2 \bar y \Gamma^{m}
(1+\Gamma_{M2})\kappa+O(y^3)
\eea 
while for supersymmetry transformations we get 
\bea 
\label{sm2} 
\delta_{\varepsilon}y&=&\varepsilon+O(y^2)\ , \cr
\delta_{\varepsilon}x^{m}&=&-\frac i2 \bar y \Gamma^{m}
\varepsilon +O(y^3). 
\eea
where the spinor $\varepsilon$ satisfies the killing spinor equation $\tilde D_{m}\varepsilon =0$.

\section{The fundamental string actions in type IIA/B}

 The F1 action in type IIA can be obtained from a double dimensional reduction of the M2-brane action (\ref{m2}), then by means of a t-duality transformation, we obtained the corresponding type IIB version. Here, we just show the final form of the actions, although long and tedious calculations were needed  in order  to obtain the type IIA expression. Also generalizations of the t-duality transformation were introduced to work out the fundamental type IIB string. It is important to say that these actions come in terms of the ``natural'' operators appearing in the supersymmetric transformation of the supergravity background fields. In the same footing, the well-known symmetries of these actions, like $\kappa$-symmetry and supersymmetry, are almost obvious from the expressions that we write here. The form of the actions is:
\bea
S_{(F)}&=&-T_{F1}\int{d^2\xi\sqrt{-\det(g)}}+\hbox{${T_{F1}\over
2}$}\int{d^2\xi\epsilon^{ij}b_{ij}} \;+ \cr
&&+iT_{F1}\int d^2\xi\sqrt{-\det(g)}\;\bar{y}\tilde P_{(-)}\Gamma^i\tilde D_iy,
\label{f}
\eea
where for the type IIA case,
\bea
y=\left(y_+\;+\;y_-\right)\;\;\hbox{with}\;\;\Gamma^{11} y_{\pm}=\pm y_{\pm}\;\;\hbox{and}\;\; P_{(-)}=\hbox{${1\over 2}$}\left(1-\hbox{${1\over 2\sqrt{-g}}$}\epsilon^{ij}\Gamma_{ij}\Gamma^{11}\right)\;,
\label{fa}
\nonumber
\eea
\bea
\tilde{D}_m &=&D^{(0)}_m+W_m\;\;,\nn
&&D^{(0)}_m = \partial_m +\frac{1}{4}\omega_{mab}\Gamma^{ab}+
\frac{1}{4\cdot2!}H_{mab}\Gamma^{ab}\Gamma^{{11}}\;, \\
&&W_m = \frac18 e^\phi \left( \frac{1}{2!} {\bf F}^{(2)}_{ab}\Gamma^{ab}\Gamma_m\Gamma^{{11}}+ \frac{1}{4!}{\bf F}^{(4)}_{abcd}\Gamma^{abcd}\Gamma_m\right)\;.\nonumber
\label{fb}
\eea
and for the type IIB case,
\bea
y=\left(
\begin{array}{cc}
y_1\\
y_2\\
\end{array}\right)\;\;\; \hbox{with} \;\;\Gamma^{11} y_{(1,2)}=y_{(1,2)}\;\; \hbox{and}\;\; P_{(-)}=\hbox{${1\over 2}$}\left(1-\hbox{${1\over 2\sqrt{-g}}$}\sigma_3\otimes\epsilon^{ij}\Gamma_{ij}\Gamma^{11}\right)\;, \nonumber
\eea
\bea
\tilde{D}_m &=&\left(
         \begin{array}{cc}
      \hat D^{(0)}_{(1)m}&0 \\
          0&\hat D^{(0)}_{(2)m}
      \end{array}\right)\;\; +\;\;
     \left(
         \begin{array}{cc}
      0&\hat W_{(2)m} \\
          \hat W_{(1)m}&0
      \end{array}\right)  \;,\nn \nn
&&\hat D^{(0)}_{(1,2)m} = \partial_m +\frac{1}{4} \omega_{mab}\Gamma^{ab}\pm \frac{1}{4\cdot 2!}H_{mab}\Gamma^{ab}\;,\\
&&\hat W_{(1,2) m} = \mp\frac18 e^\phi \left( {\bf F}^{(1)}_a\Gamma^a \pm \frac{1}{3!} {\bf F}^{(3)}_{abc}\Gamma^{abc}+
\frac{1}{2\cdot 5!}{\bf F}^{(5)}_{abcde}\Gamma^{abcde}\right)\Gamma_m \;, \nonumber
\eea
where in the expressions for $\hat D^{(0)}$ and $\hat W$, $\pm$ correspond to the label $(1,2)$. Also in all the above, ($m,n..$) are 10D space-time indices, $g$ is the 10D metric, $H=db$ and ${\bf F}^{(n)}$ are the RR field strength. Note that $\tilde D_m$ is precisely the operator appearing in the supersymmetry variation of the 10D gravitino, i.e. $\delta_\epsilon \psi_m=\tilde D_m \epsilon$ (see the appendix on \cite{ms,mms1,mms2} for the explicit expressions, conventions and definitions).

The corresponding $\kappa$-symmetry transformations up to second order in $y$ are
\bea
\delta_\kappa y^\alpha = (1+\Gamma_F)\kappa \;\;\;,\;\;\;
\delta_\kappa x^m= {i\over 2}\bar y \Gamma^m (1+\Gamma_F)\kappa\;\;\;,\;\;\;
\delta_\kappa A= \delta_\kappa x^m\partial_m A\;,
\label{akappa}
\eea
where $\Gamma_F=\hbox{${1\over 2\sqrt{-g}}$}\epsilon^{ij}\Gamma_{ij}\Gamma^{11}$ and $A$ is a general field of the supergravity background. The supersymmetry transformations (again up to second order in $y$) are
\bea
\delta_\epsilon y =\epsilon \;\;\;,\;\;\;
\delta_\epsilon x^m= -{i\over 2}\bar y \Gamma^m \kappa \;\;\;,\;\;\;
\delta_\epsilon A= \delta_\epsilon x^m\partial_m A\;,
\label{asusy}
\eea
where again $A$ is a general field of the supergravity background and the fermion $\epsilon$ is actually a killing spinor of the bosonic background, i.e. $\delta_\epsilon\psi_m=\delta_\epsilon\lambda=0$, where $\psi_m$ and $\lambda$ are the gravitino and the dilatino respectively.

\section{D-brane actions in type IIA/B}

To obtain the general form of the D-brane actions, we performed first a single dimensional reduction from the M2-brane action to 10D to obtain the D2-brane action of type IIA theory. Then by means of t-duality we recovered all the other D-brane actions. Although conceptually the program is clear, it is far from straightforward since the calculation is plague with technical problems and subtleties like for example, the generalization of t-duality rules for the worldvolume fermions and its compatibility with already existent rules for supergravity like those given by Hassan \cite{has}.

The resulting actions are
\bea
S_{Dp}&=&S^{(0)}_{Dp}+S^{(2)}_{Dp}+O(y^{4}),\nn
S_{Dp}^{(0)}&=& -T_{Dp}\int d^3\xi e^{-\phi}\sqrt{-(g+{\cal
F})}+T_{Dp}\int C\; e^{-{\cal F}}\,, \cr 
S^{(2)}_{Dp}&=&\frac
{iT_{Dp}}{2}\int d^3\xi e^{-\phi}\sqrt{-(g+{\cal
    F})}\bar y(1-\tilde\Gamma_{D_p})(\Gamma^iD_i-\Delta+L_p)y .
\label{dpactions2}
\eea
with
\bea
\label{Ggeneral}
\tilde\Gamma_{D(2n)}=\hbox{$\frac{1}{\sqrt{-(g+{\cal F})}}$}\sum_{q+r=n}\hbox{$\frac{\epsilon^{i_1\ldots i_{2q}j_1\ldots j_{2r+1}}}{q!2^q(2r+1)!}$}{\cal F}_{i_1i_2}\cdots{\cal F}_{i_{2q-1}i_{2q}}\Gamma_{j_1\ldots j_{2r+1}}(\Gphi)^{r+1}\ ,\cr
\tilde\Gamma_{D(2n+1)}=\hbox{$\frac{-i\sigma_2}{\sqrt{-(g+{\cal F})}}$}\sum_{q+r=n+1}\hbox{$\frac{\epsilon^{i_1\ldots i_{2q}j_1\ldots j_{2r}}}{q!2^q(2r)!}$}{\cal F}_{i_1i_2}\cdots{\cal F}_{i_{2q-1}i_{2q}}\Gamma_{j_1\ldots j_{2r}}(\hGphi)^{r}.
\eea
\bea
L_{2n+1}&&=\sum_{q\geq 1,q+r=n+1}\hbox{$\frac{\epsilon^{i_1\ldots i_{2q}j_1\ldots j_{2r+1}}(-i\sigma_2)(\hGphi)^{r}}{q!2^q(2r+1)!\sqrt{-(g+{\cal F})}}{\cal F}_{i_1i_2}\cdots{\cal F}_{i_{2q-1}i_{2q}}\Gamma_{j_1\ldots j_{2r+1}}{}^k \hat D_k$} \ ,\cr
L_{2n}=&&\sum_{q\geq 1,q+r=n}\hbox{$\frac{\epsilon^{i_1\ldots i_{2q}j_1\ldots
    j_{2r+1}}(-\Gphi)^{r+1}}{q!2^q(2r+1)!\sqrt{-(g+{\cal F})}}$}\hbox{${\cal
  F}_{i_1i_2}\cdots{\cal F}_{i_{2q-1}i_{2q}}\Gamma_{j_1\ldots j_{2r+1}}{}^k D_k$}\ ,
\eea
where for the type IIA case
\begin{eqnarray}\label{operatorsIIA}
D_m &=& D^{(0)}_m+W_m \cr
\Delta &=& \Delta^{(1)}+\Delta^{(2)}\ ,
\end{eqnarray}
with
\begin{eqnarray}
D^{(0)}_m &=& \partial_m +\frac{1}{4} \omega_{mab}\Gamma^{ab}+\frac{1}{4\cdot 2!}H_{mab}\Gamma^{ab}\Gamma^{\ul{\varphi}} \cr
W_m &=& \frac18 e^\phi \left( \frac{1}{2!} {\bf F}^{(2)}_{ab}\Gamma^{ab}\Gamma_m\Gamma^{\ul{\varphi}}+
\frac{1}{4!}{\bf F}^{(4)}_{abcd}\Gamma^{abcd}\Gamma_m\right)\cr
\Delta^{(1)} &=& \frac12 \left( \Gamma^m \partial_m\phi +\frac{1}{2\cdot 3!}H_{abc}\Gamma^{abc}\Gamma^{\ul{\varphi}}\right)\cr
\Delta^{(2)}&=& \frac{1}{8} e^\phi \left( \frac{3}{2!} {\bf F}^{(2)}_{ab}\Gamma^{ab}\Gamma^{\ul{\varphi}}+
\frac{1}{4!} {\bf F}^{(4)}_{abcd}\Gamma^{abcd}\right)\ .
\end{eqnarray}
and for the type IIB case
\bea
\label{operatorsIIB}
\hat D_m &=& \hat D^{(0)}_m+\sigma_1 \otimes \hat W_m \cr
\hat \Delta &=& \hat\Delta^{(1)}+\sigma_1\otimes\hat\Delta^{(2)}\ . 
\eea
with
\begin{eqnarray}
\hat D^{(0)}_{(1,2)m} &=& \partial_m +\frac{1}{4} \omega_{mab}\Gamma^{ab}\pm \frac{1}{4\cdot 2!}H_{mab}\Gamma^{ab}\cr
\hat W_{(1,2) m} &=& \frac18 e^\phi \left(\mp {\bf F}^{(1)}_a\Gamma^a - \frac{1}{3!} {\bf F}^{(3)}_{abc}\Gamma^{abc}\mp
\frac{1}{2\cdot 5!}{\bf F}^{(5)}_{abcde}\Gamma^{abcde}\right)\Gamma_m\cr
\hat\Delta^{(1)}_{(1,2)} &=& \frac12 \left( \Gamma^m \partial_m\phi \pm\frac{1}{2\cdot 3!}H_{abc}\Gamma^{abc}\right)\cr
\hat\Delta^{(2)}_{(1,2)}&=& \frac{1}{2} e^\phi \left( \pm  {\bf F}^{(1)}_{a}\Gamma^{a}+
\frac{1}{2\cdot 3!} {\bf F}^{(3)}_{abc}\Gamma^{abc}\right)\ .
\end{eqnarray}
where $\Gamma^{11}=\Gamma^{\underline\phi}$ and $\hat\Gamma^{\underline\phi}=\sigma_3\otimes \Gamma^{\underline\phi}.$

\vspace{1cm}
{\bf Acknowledgments}\\

We thank M. Grisaru, R. Myers and D. Zanon for useful discussions. L. Martucci and P. J. Silva were partially supported by INFN, MURST and by the European Commission RTN program HPRN-CT-2000-00131, in association with the University of Torino. D. Marolf and P. J. Silva were supported in part by NSF grant PHY00-98747 and by funds from Syracuse University.
\\


\end{document}